\documentclass[prd,11pt,nofootinbib,preprintnumbers,final]{revtex4}
  \usepackage{bm}
   \usepackage{amsmath,amsthm}
    \usepackage{amssymb}
     \usepackage{pifont}
\usepackage{graphicx}

\renewcommand{\(}{\left(}
\renewcommand{\)}{\right)}
\renewcommand{\[}{\left[}
\renewcommand{\]}{\right]}
\newcommand{\nn}{\nonumber}

\begin{document}

\title{Chiral expansion of nucleon PDF at $x\sim m_\pi/M_N$}

\author{A.Moiseeva}
\affiliation{Institut f\"ur Theoretische  Physik II, Ruhr\\University Bochum, 44780 Bochum, Germany}
\email{alenasmail@gmail.com}
\author{A.A.Vladimirov}
\affiliation{Department of Astronomy and Theoretical Physics, Lund University,\\ S\"olvegatan 14A, S 223 62 Lund, Sweden}
\email{vladimirov.aleksey@gmail.com}

\begin{abstract}
Based on the chiral perturbation theory, we investigate the low-energy dynamics of nucleon parton distributions. We show that in different
regions of the momentum fraction $x$ the chiral expansion  is significantly different. For nucleon parton distributions these regions are
characterized by $x\sim1$, $x\sim m_\pi/M_N$ and $x\sim (m_\pi/M_N)^2$. We derive extended counting rules for each region and obtain
model-independent results for the nucleon parton distributions down to $x\gtrsim m^2_\pi/M^2_N \approx 10^{-2}$.
\end{abstract}

\preprint{LU TP 13-39} \preprint{Nov. 2013 }

\maketitle

\section{Introduction}
\label{intro}

The investigation of the pion contribution to nucleon parton distribution functions (PDFs) started already in the early 70's \cite{Drell:1969wd}. Roughly speaking,
one can single out two utmost approaches. The first approach is based on the convolution model and on the interpretation of the pion cloud distribution
as the amplitude of the Sullivan process, for details see \cite{Speth:1996pz}. The second one is based on the straightforward application of
low-energy effective theories, such as chiral perturbation theory (ChPT), to PDFs, see e.g.~\cite{Arndt:2001ye}. Both of these
approaches have their own advantages and disadvantages: the convolution model provides a simple and demonstrative interpretation, whereas the effective
theory approach is based on a systematical expansion.
In spite of superficial similarity of the approaches some of the results for the meson cloud contributions are in
contradiction, for a recent comparison see \cite{Burkardt:2012hk}.

The application of ChPT  to  pions shows itself to be very efficient and describes from the model-independent field theory-based point of view
some well-known effects, such as, the increasing of the pion size in the chiral limit \cite{Perevalova:2011qi}. On the other hand the
application of the low-energy effective theory to the non-local operators, appearing in the definition of parton distributions, is not
straightforward. This is caused by the necessity  to resum  higher order terms in the chiral expansion \cite{Kivel:2007jj}. Recently such an
analysis has been performed for the nucleon case \cite{Moiseeva:2012zi}. It has been shown that in the nucleon case the situation is more
distinct, i.e., there are several regions in $x$ with significantly different structure of the chiral expansion, which has been earlier
recognized in model considerations, see e.g.\cite{Strikman:2009bd}. In the following we shortly explain the uses of the meson-nucleon ChPT
framework for the evaluation of nucleon parton distributions. The details and results of our calculation are/will be given in
\cite{Moiseeva:2012zi,Moiseeva_PhD,TOBE}.

\begin{figure}
\centering
  \includegraphics[width=0.7\textwidth]{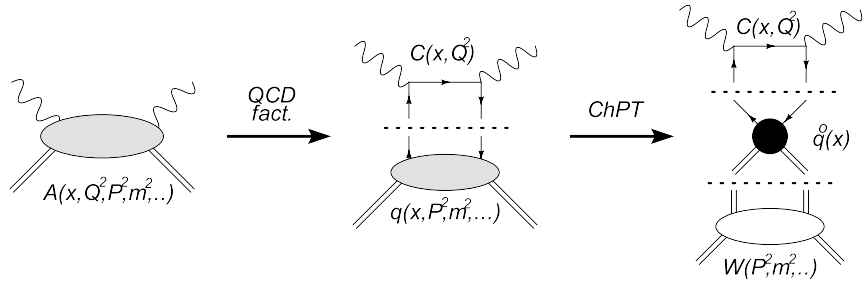}
\caption{Diagrammatical illustration of a hard scattering process. The coefficient function $C$ is computable within pertubative QCD, whereas
the coefficient function $W$ is computable within ChPT. The horizontal lines represent the Mellin convolution and the black blob is the parton
distribution in the chiral limit.}  \label{fig:0}
\end{figure}

\section{Effective operator for nucleon parton distributions}
\label{sec:1}

Generalized parton distributions (GPDs) are very intricate and rich objects. To point out the framework, we consider
for simplicity only non-skewed GPDs ($\Delta^+=0$) with nonzero $\Delta^2$-dependance, which reduce at $\Delta^2=0$ to nucleon PDFs, where we
restrict us to the isovector combination. Expressions for $\Delta^+\neq0$, as well as, for other flavor
combinations can be found in \cite{Moiseeva_PhD}. These
parton densities are defined via the nucleon matrix elements of the light-cone quark operators
\begin{eqnarray}\label{def}
\int \frac{d\lambda}{2\pi}e^{-ix\lambda P^+}\langle
p|O(\lambda)|p'\rangle\bigg|_{\Delta_+=0}=\frac{1}{P^+}\bar
u\(\gamma^+
q^{u-d}(x,\Delta^2)+\frac{i\sigma^{+\nu}\Delta_\nu}{2M}E^{u-d}(x,\Delta^2)\)u,
\end{eqnarray}
where $O(\lambda)$ is the unpolarized (or vector) quark operator of twist-two, $M$ is the nucleon mass, and $\Delta=p'-p$.

To calculate the matrix element  (\ref{def}) in an effective field theory, we have to find an effective operator in terms of hadronic degrees of
freedom that possesses the symmetry properties as the QCD operator $O(\lambda)$. This procedure can be understood as some kind of low-energy
factorization, which separates the very low-energy dynamics of large distance interactions of a hadron with its meson cloud (which is governed
by the spontaneous chiral symmetry braking) from the unknown dynamics of the hadron core. Visual representation of such a double factorization
is presented in fig.\ref{fig:0}. The unknown part $\mathring q(x)$, representing the hadron core dynamics, enters via a generating function in
the effective operator. Since the construction of the effective operator is only restricted by the quantum numbers of the QCD operator, the
amount of possible effective operators is infinite. However, their number can be restricted by choosing a proper counting hierarchy. Note that
the effective operator has indefinite twist since the effective degrees of freedom are of indefinite twist by themselves.

In order to calculate the chiral corrections to parton distributions, one should first find the counting rules for all dimensional quantities. In the
meson-baryon ChPT one has the following counting rules
\begin{eqnarray}\label{stnd_cr}
\partial_\mu \pi \sim m,~~~~~~~~\partial_\mu N\sim M,
\end{eqnarray}
where the pion mass $m \ll M$. The pion mass is the small parameter of the chiral expansion $\frac{m}{4\pi
F_\pi}=a_\chi\ll 1$, whereas, the nucleon mass violates the low-energy expansion: $\frac{M}{4\pi F_\pi}\sim 1$. There are several methods to
bypass the violation of the chiral expansion by nucleon mass, such as, the heavy baryon theory \cite{Jenkins:1990jv}, the extended on-mass-shell
(EOMS) scheme \cite{Fuchs:2003qc}, and several others. However, for the consideration of a nonlocal  operator one should go beyond the standard
power counting rules of meson-baryon ChPT. The reason is that the nonlocal operator has its own intrinsic dimensional scale: the light-cone
separation $\lambda$. The chiral counting for $\lambda$ should be defined additionally.

At all that, the situation is significantly different for pion and nucleon parton distributions. The origin of the difference is the
counting rules for the derivatives of pion and nucleon fields (\ref{stnd_cr}). Let us demonstrate this fact explicitly. First of all, in order
to apply the counting rules, we have to expand the matrix element of the light-cone operator in the set of local operators,
\begin{eqnarray}\label{deriv_exp}
O(\lambda)=O^{(0)}+\lambda O^{(1)}+\lambda^2 O^{(2)}+...~,
\end{eqnarray}
where $O^{(n)}\sim \partial_+^n O(0)$.

Let us suppose now that the operator (\ref{deriv_exp}) contains only pion fields and that only a pion is present in the
in/out-state. Then the chiral expansion for every individual local operator $O^{(n)}$ starts from $a_\chi^n$:
\begin{eqnarray}
\langle \pi|O_\pi(\lambda)|\pi\rangle=\[q^{(0,0)}+a_\chi q^{(0,1)}+...\]+a_\lambda \[a_\chi q^{(1,1)}+a^2_\chi q^{(1,2)}+...\]+
a_\lambda^2\[a_\chi^2 q^{(2,2)}+...\]+...~,
\end{eqnarray}
where $a_\lambda=(4\pi F_\pi)\lambda$. One can see that if $a_\lambda\sim a_\chi$ or $a_\lambda\sim 1$ only the first few terms contribute to
the expansion at level $\mathcal{O}(a_\chi)$.  Such a picture corresponds to rather small light-cone separation, $\lambda\sim (4\pi
F_\pi)^{-1}$. In the regime $a_\lambda\sim a_\chi^{-1}$, which implies $\lambda\sim m^{-1}$, the series reorganizes:
\begin{eqnarray}
\langle \pi|O_\pi(\lambda)|\pi\rangle=\[q^{(0,0)}+a_\chi a_\lambda q^{(1,1)}+(a_\chi a_\lambda)^2 q^{(2,2)}+...\]+a_\chi \[q^{(0,1)}+a_\chi
a_\lambda q^{(1,2)}+...\]+...~,
\end{eqnarray}
where all terms in the brackets are of the same order. For large $\lambda$, the higher order contributions of the expansion should be taken into
account. However, in the definition of the parton distribution (\ref{def}) all possible $\lambda$'s contribute, except for very large $\lambda$ at
which the Fourier exponent starts to oscillate. The effective region of integration is $0<\lambda \lesssim(x p_+)^{-1}$. Therefore, in order to
obtain the correct chiral expansion in, say $x\sim 1$, one should take $a_\lambda\sim a_\chi^{-1}$ at least (since $p_+\sim m$). The detailed
discussion can be found in \cite{Kivel:2002ia}. For lower $x$ the higher order terms should be taken into account \cite{Kivel:2007jj}.

For the nucleon operator, or for the pion operator in the nucleon
brackets, the structure of the chiral expansion is different. The
point is that there is a possibility to get the nucleon mass scale
via derivatives acting on the nucleon field, and therefore
\begin{eqnarray}\label{N_exp}
\langle N|O_N(\lambda)|N\rangle=\[q^{(0,0)}+a_\chi q^{(0,1)}+...\]+a_\lambda \[q^{(1,0)}+a_\chi q^{(1,1)}+...\]+
a_\lambda^2\[q^{(2,0)}+...\]+...~.
\end{eqnarray}
One can see that at $a_\lambda\sim 1$(or $\lambda \sim M^{-1}$)  the
expansion (\ref{N_exp}) contains infinitely many  terms of
$\mathcal{O}(a_\chi)$. Integration over the region
$0<\lambda\lesssim M^{-1}$ corresponds to $x\sim1$.
This regime has been
considered in \cite{Chen:2001eg,Arndt:2001ye,Diehl:2006js} and in many other articles. At such small $\lambda$ the operator is almost local, and the result of calculation leads to the generally  incorrect expression $q(x,\Delta^2)= q(x)F(\Delta^2)$,
where $F$ is the corresponding form factor. The first significant reorganization of the series (\ref{N_exp}) takes place at $a_\lambda \sim
a_\chi^{-1}$. This allows us to obtain corrections to parton distributions down to $x\sim \frac{m}{M}=\alpha$. The next significant
reorganization takes place at $a_\lambda\sim a_\chi^{-2}$, which corresponds to $x\sim \alpha^2$. More details of the chiral
expansion analysis for nucleon parton distributions are given in \cite{Moiseeva:2012zi}.

It is very inconvenient to deal with the parameter $\lambda$ in a straightforward manner, moreover, due to the $\lambda$-integration
we loose the guidance of the counting rules. In order to bypass these difficulties, we suggest to transfer the counting rules of $\lambda$ to
the light-cone vector $n_\mu$, which  accompanies the parameter $\lambda$ in the definition of the parton
distribution. Also we suggest  to work directly with nonlocal operator (i.e., without expansion in $\lambda$). Note that one can rescale $n_\mu$ without damaging the operator properties. Thus, we assume that $\lambda\sim 1$, whereas
$n_\mu$ changes its counting depending on $x$. In the $x$-region, interesting for us, the counting rule for $n_\mu$ reads
\begin{eqnarray}\label{counting}
n_\mu\sim \frac{1}{m}~~~~~\text{for}~~x > \frac{m^2}{M^2}=\alpha^2.
\end{eqnarray}
 The assumption $\lambda\sim 1$ prevents us from the expansion of the nonlocal operator. Therefore,  all loop calculations are performed with nonlocal vertices.

Employing the new counting rules,  we have constructed the operator suitable for the description of GPDs and  PDFs in the range down to $x\sim
\alpha$ for the vector and axial-vector cases in both the isovector and isoscalar sectors, which arise altogether from four different generating functions. Only two of them appear at the tree level and in the chiral limit they have the meaning of the corresponding parton distributions. The remaining two functions appear at one loop level and they have no simple interpretation. The isovector vector operator is a typical representative,
\begin{eqnarray}\label{O1}
O^a(\lambda)=\int_{-1}^1d\beta~ \bar N\(-\frac{\beta
\lambda}{2}\)\gamma^+\Big[\mathring q(\beta)
t^a_++q_2(\beta)\gamma^5 t_-^a+\frac{\mathring
q(\beta)-q_3(\beta)}{4}\tilde
t_+^a~~~~~~~~~~~~~\\\nn+\frac{\Delta\mathring
q(\beta)-q_2(\beta)}{4}\gamma^5\tilde
t_-^a\Big]N\(\frac{\beta\lambda }{2}\),
\end{eqnarray}
where $\mathring q(\beta)$ and $\Delta \mathring q(\beta)$ are the isovector combinations of PDFs in the chiral limit, $q_2(\beta)$ and
$q_3(\beta)$ are additional generating functions. The pion field combinations are
$$
t^a_{\pm}=\frac{1}{2}\(u^\dagger\(-\frac{\beta\lambda}{2}\)\tau^a u\(\frac{\beta\lambda}{2}\)\pm u\(-\frac{\beta\lambda}{2}\)\tau^a
u\(\frac{\beta\lambda}{2}\)\),
$$
$$
\tilde
t^a_{\pm}=\frac{1}{2}\(u^\dagger\(-\frac{\beta\lambda}{2}\)\[\tau^a,U\(\frac{\beta\lambda}{2}\)\]
u^\dagger\(\frac{\beta\lambda}{2}\)\pm
u\(-\frac{\beta\lambda}{2}\)\[\tau^a,
U^\dagger\(\frac{\beta\lambda}{2}\)\]
u\(\frac{\beta\lambda}{2}\)+...\),
$$
where $u^2=U=\exp(i\pi^a \tau^a/2 F_\pi)$, and the dots denote the terms with commutators at the point $\(-\beta \lambda/2\)$. The axial-vector
operator contains the same generating functions. Additionally, there is an operator which contains only the pion fields. It was derived in
\cite{Kivel:2002ia} and reads
\begin{eqnarray}\label{O2}
O^a_{\pi}(\lambda)=\frac{-i F_\pi^2}{4}\int_{-1}^1d\beta \mathring
Q(\beta) \text{Tr}\[\tau^a
\(U^\dagger\(-\frac{\beta\lambda}{2}\)\overset{\text{\scriptsize$\leftrightarrow$}}{\partial}_+U\(\frac{\beta\lambda}{2}\)
+U\(-\frac{\beta\lambda}{2}\)\overset{\text{\scriptsize$\leftrightarrow$}}{\partial}_+U^\dagger\(\frac{\beta\lambda}{2}\)\)\],
\end{eqnarray}
where $\mathring Q(x)$ is  the isovector pion PDF in the chiral limit. The complete expressions for the operators, and they normalizations can
be found in \cite{Moiseeva_PhD,TOBE}.

\section{Chiral structure of the nucleon} \label{sec:2}
\begin{figure}
\centering
  \includegraphics[width=0.49\textwidth]{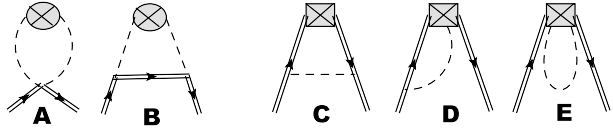}
\caption{Diagrams relevant for the leading chiral correction to the
unpolarized (vector) nucleon parton distribution. The crossed circle denotes the
pure pion operator (\ref{O2}), the crossed box denotes the
pion-nucleon operator (\ref{O1}).} \label{fig:1}
\end{figure}

One of the main consequence of the different counting rules, applied to different regions in $x$, is that the resulting expression for the leading chiral
correction is non-linear in $a_\chi^2$ rather than linear. The situation should be understood in the following way. On
one hand, there is an ``exact" (containing all orders of the perturbative expansion) expression $q(x)=\sum_{n=0}^\infty a_\chi^n q^{(n)}(x)$. On
the other hand, there is a truncated  expression which contains, say, the leading term and the next-to-leading term: $q_{t}(x)=q^{(0)}(x)+a_\chi^2
q^{(1)}(x)$. The difference between these two functions, $q(x)-q_t(x)$, is not necessarily $\mathcal{O}(a_\chi^3)$, but depending on $x$ it can
be $\mathcal{O}(a_\chi^3)$ or $\mathcal{O}(a_\chi^0)$. Resumming particular higher order contributions, we obtain the expression which differs
from the unknown ``true" expression by $\mathcal{O}(a_\chi^3)$ contributions.

Using the operators (\ref{O1}-\ref{O2}) and the counting rules (\ref{counting}), one can evaluate the leading nonanalytical contribution to the
nucleon parton distributions. The analytical part cannot be evaluated that easy because it contains a large set of new generating functions. The
diagram representation of the leading contribution is shown in fig.~\ref{fig:1}. The  nucleon parton distribution is conveniently
presented in the following form
\begin{eqnarray}\label{q(Delta)}
q(x,\Delta^2)=\mathring q(x)+\frac{M^2}{(4\pi F_\pi)^2}\int_{-1}^1 \frac{d\beta}{|\beta|}\theta\(0<\frac{x}{\beta}<1\)\Big[\mathring
q\(\frac{x}{\beta}\) C(\beta,\Delta^2) ~~~~~~~~~~~~~~~~\\\nn+\(\Delta \mathring q\(\frac{x}{\beta}\)-\frac{\mathring
q_2\(\frac{x}{\beta}\)}{2}\)\Delta C(\beta,\Delta^2)+\mathring Q\(\frac{x}{\beta}\)C_\pi(\beta,\Delta^2)\Big],
\end{eqnarray}
where
\begin{eqnarray}
C(x,\Delta^2)&=&-\(1+\frac{5g_a^2}{2}\)\delta(\bar x)\alpha^2\ln\alpha^2+g_a^2\int_0^{\bar x} d\eta\frac{x \alpha^2-\frac{\Delta^2}{M^2}(\bar
x-\eta(1+x^2))}{\bar x^2+\alpha^2 x -\frac{\Delta^2}{M^2}\eta(\bar x-\eta)},
\\
\Delta C(x,\Delta^2)&=&g_a\delta(\bar x)\alpha^2\ln\alpha^2+2g_a \bar x \ln\(1+\frac{\alpha^2 x}{\bar x^2}\),
\\
C_\pi(x,\Delta^2)&=&(1-g_a^2)\delta(x)\int_0^1d\eta \(\alpha^2-\eta\bar \eta \frac{\Delta^2}{M^2}\)\ln  \(\alpha^2-\eta\bar \eta
\frac{\Delta^2}{M^2}\)
\\\nn&&~~~~~~~~~~-4g_a^2 x\ln\(1+\frac{\alpha^2 \bar x}{x^2}\)
+4g_a^2\int_0^{\bar x} d\eta \frac{x\alpha^2-x\eta \frac{\Delta^2}{M^2}}{x^2+\bar x \alpha^2-\frac{\Delta^2}{M^2}x\(\bar x-\eta\)},
\end{eqnarray}
with the axial-vector coupling constant $g_a\approx 1.27$, $\alpha=\frac{m}{M}\approx 0.15$, and $\bar x = 1-x$. The evaluation has been done
within the EOMS renormalization scheme. The structure of our result typically appear in a dispersive framework, see e.g.
\cite{Alberg:2012wr,Granados:2013moa}, but never occurred in straightforward ChPT calculations \cite{Arndt:2001ye,Chen:2001eg}.

\begin{figure}
\centering
  \includegraphics[width=0.49\textwidth]{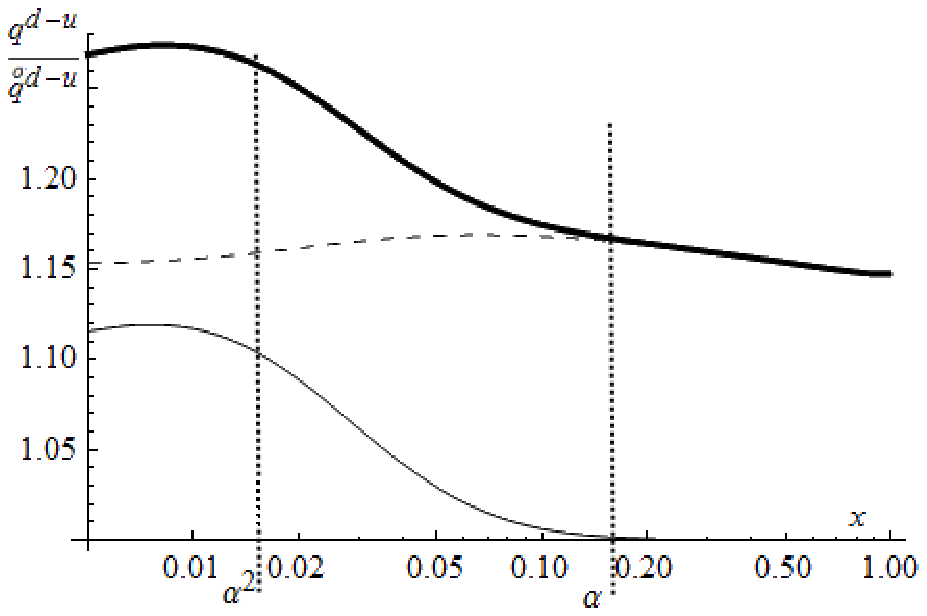}
    \includegraphics[width=0.49\textwidth]{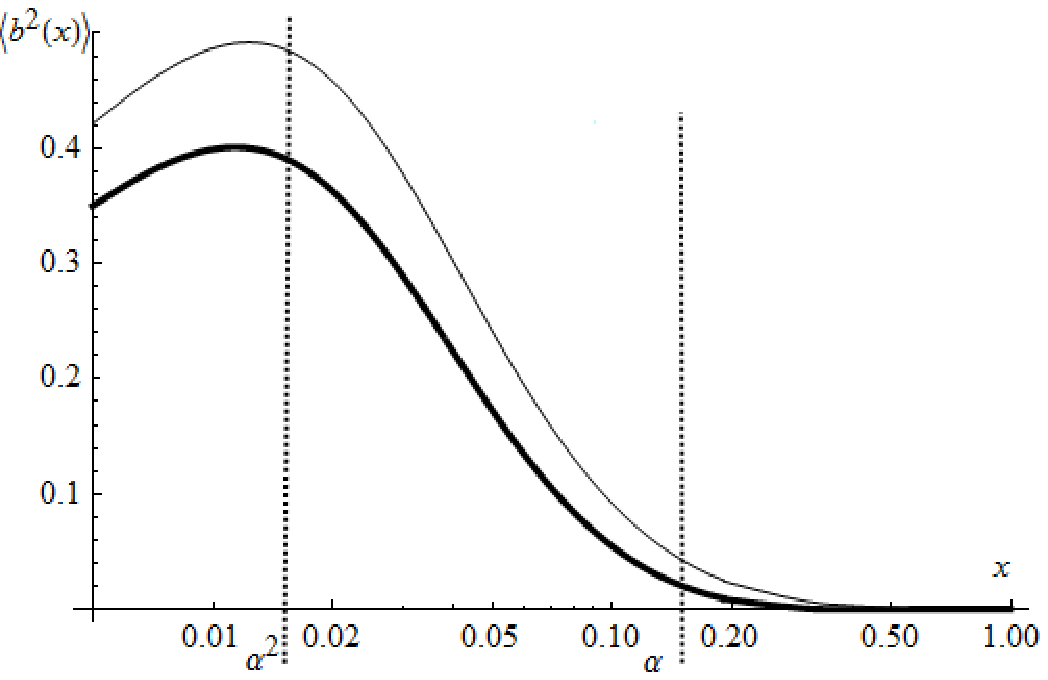}
\caption{Left panel: relative chiral corrections to the PDF (thick curve), without pion operator (dashed curve), and  only pion operator (thin
curve). Right panel: The transverse  size of the nucleon (thick curve) and only pion operator contribution (thin curve).}
\label{fig:2}       
\end{figure}

Indeed, one can  see that in the region $x\sim 1$  our expressions  can be straightforwardly expanded in $\alpha$. The first terms of the
expansion coincide with the results obtained by means of Mellin moments, see e.g. \cite{Diehl:2006js}. The region $x\sim \alpha$ is not
described by the first terms of the $\alpha$-expansion, although it is still of order $\mathcal{O}(\alpha^2)$. In the region $x\sim\alpha^2$
 one cannot expand the functions  $C,~\Delta C$ and $C_\pi$ in $\alpha$, since they are altogether of order
$\alpha^2$. These results are in agreement with \cite{Burkardt:2012hk} and can be compared in parts with dispersion analysis in
\cite{Granados:2013moa}.

Having at hand the model-independent result for the low-energy parameter behavior of the nucleon parton distributions one can consider various
interesting aspects parton dynamics, such as, the role of the pion cloud in the nucleon, the size of the nucleon, the large-distance behavior of
the nucleon, and many others. A brief inspection of expressions (\ref{q(Delta)}) shows the significant dominance of the pion operator in the
region $x <\alpha$. This is also visualized in the left panel of fig.~\ref{fig:2}, where we
show the relative chiral corrections for a standard PDF parametrization (taken as PDF in the chiral limit%
\footnote{A more realistic treatment requires to take a PDF that is calculated in some model with massless quarks, e.g., in the light-front
formulation of QCD.}). One can see that the main contribution comes from the pion operator and that it this unevenly grows in the region
$\alpha^2 \lesssim x \lesssim \alpha$. The behavior demonstrated in fig.~\ref{fig:2} is universal, and holds for all quark and also antiquark
PDFs. Recently,  experimental data have been confronted with the description of the pion cloud model by means of the Sullivan process
\cite{Traini:2013zqa}. The main disagreement of data and the model prediction takes place in the region $x\sim\alpha$. Possibly, this
disagreement can be reduced in our approach.

A more clear and model-independent result follows from the $\Delta^2$-dependance. The point is that the parton distribution in the chiral limit
can be expressed via the standard PDF by solving the expression (\ref{q(Delta)}) at $\Delta^2=0$. Moreover, the transverse size of the nucleon,
defined as $\langle b^2(x)\rangle=4 d\ln q(x,\Delta^2)/d\Delta^2$ at $\Delta^2=0$, is insensible to the difference between $q(x)$ and $\mathring
q(x)$ and also does not contain the unknown functions $\mathring q_2(x)$ and $\mathring q_3(x)$. Therefore, one can very precisely and in a
model-independent way obtain an expression for the transverse size of the nucleon for the region $x\gtrsim \alpha^2$. In the right panel of
fig.~\ref{fig:2} we show our result for the transverse size vs.~$x$. Technical details of our approach and qualitative estimates will be
presented in \cite{TOBE}.

\textbf{Acknowledgements~}{\small A.M. is supported in part by DFG (SFB/TR 16, ``Subnuclear Structure of Matter'') and by the European
Community-Research Infrastructure Integrating Activity ``Study of Strongly Interacting Matter'' (acronym HadronPhysics3, Grant Agreement n.
283286) under the Seventh Framework Programme of EU. A.V. thanks the organize committee of Light-Cone 2013 for support and also A.V. is
supported in part by the European Community-Research Infrastructure Integrating Activity Study of Strongly Interacting Matter (HadronPhysics3,
Grant Agreement No. 28 3286) and the Swedish Research Council grants 621-2011-5080 and 621-2010-3326.}

\end{document}